\documentclass[runningheads]{llncs}
\usepackage{graphicx}
\usepackage{tikz}
\usepackage{comment}
\usepackage{amsmath,amssymb} % define this before the line numbering.
\usepackage{color}

\usepackage{hyperref}
\usepackage{caption}
\usepackage{subcaption}
\usepackage{extarrows}
\usepackage{threeparttable}
\usepackage{multirow}
\usepackage{booktabs}

\usepackage{enumitem}
\usepackage{pifont}

\newcommand{\chkYES}{\rlap{\raisebox{0.3ex}{\hspace{0.4ex}\tiny \ding{52}}}$\square$}
\newcommand{\chkNO}{\rlap{\raisebox{0.3ex}{\hspace{0.4ex}\scriptsize \ding{56}}}$\square$}

\newcommand{\I}{\textbf{I}}

\newcommand{\F}{\mathcal{F}}

\newcommand{\A}{\mathcal{A}}

\newcommand{\G}{\mathcal{G}}
\newcommand{\PP}{\mathcal{P}}

\newcommand{\R}{\mathcal{R}}
\newcommand{\N}{\mathcal{N}}
\newcommand{\X}{\mathcal{X}}

\newcommand{\eS}{\mathcal{S}}

\newcommand{\eR}{\mathbb{R}}

\newcommand{\emm}{\textbf{\emph{\emph{m}}}}
\newcommand{\ez}{\textbf{\emph{\emph{z}}}}

\newcommand{\ey}{\textbf{\emph{\emph{y}}}}

\newcommand{\ex}{\textbf{\emph{\emph{x}}}}
\newcommand{\eX}{\textbf{\emph{\emph{X}}}}
\newcommand{\eT}{\textbf{\emph{\emph{T}}}}

\newcommand{\eH}{\textbf{\emph{\emph{H}}}}
\newcommand{\eHT}{\textbf{\emph{\emph{H}}}^{\dag}}

\newcommand{\ieB}{\emph{i.e. }}
\newcommand{\egB}{\emph{e.g. }}

\begin{document}
% \linenumbers
\pagestyle{headings}
\mainmatter
\def\ECCVSubNumber{6272}  % Insert your submission number here

\title{Deep Decomposition Learning for Inverse Imaging Problems} % Replace with your title

\begin{comment}
\titlerunning{ECCV-20 submission ID \ECCVSubNumber}
\authorrunning{ECCV-20 submission ID \ECCVSubNumber}
\author{Anonymous ECCV submission}
\institute{Paper ID \ECCVSubNumber}
\end{comment}
%******************

%\begin{comment}
\titlerunning{Deep Decomposition Learning for Inverse Imaging Problems}
%\author{
%Dongdong Chen\orcidID{0000-0002-7016-9288} % index{Chen, Dongdong}
%\and
%Mike E. Davies % index{Golbabaee, Mohammad}
%}

\author{
Dongdong Chen % index{Chen, Dongdong}
\and
Mike E. Davies % index{Golbabaee, Mohammad}
}

\authorrunning{D. Chen et al.}
\institute{School of Engineering, University of Edinburgh, UK\\
% \url{https://github.com/edongdongchen/DDN}\\
\email{\{d.chen, mike.davies\}@ed.ac.uk}}

\maketitle

\begin{abstract}
Deep learning is emerging as a new paradigm for solving inverse imaging problems. However, the deep learning methods often lack the assurance of traditional physics-based methods due to the lack of physical information considerations in neural network training and deploying. The appropriate supervision and explicit calibration by the information of the physic model can enhance the neural network learning and its practical performance. In this paper, inspired by the geometry that data can be decomposed by two components from the null-space of the forward operator and the range space of its pseudo-inverse, we train neural networks to learn the two components and therefore learn the decomposition, \ieB  we explicitly reformulate the neural network layers as learning range-nullspace decomposition functions with reference to the layer inputs, instead of learning unreferenced functions. We empirically show that the proposed framework demonstrates superior performance over recent deep residual learning, unrolled learning and nullspace learning on tasks including compressive sensing medical imaging and natural image super-resolution. Our code is available at \url{https://github.com/edongdongchen/DDN}.
%\keywords{Decomposition learning, inverse problems, image reconstruction, deep learning}
\keywords{decomposition learning, physics, inverse problems}
\end{abstract}

\section{Introduction}
\label{sec:intro}
We consider a linear inverse problem  of the form :
\begin{equation}\label{eqs:bases}
    \ey_{\epsilon} = \eH\ex + \epsilon,
\end{equation}
where the goal is to recover the unknown signal $\ex\in \eR^D$ from the noisy measurement $\ey_{\epsilon}\in \eR^d$ with typical dimension $D\gg d$, and $\eH:\eR^D\rightarrow \eR^d$ is the forward operator which models the response of the acquisition device or reconstruction system, while $\epsilon \in \eR^d$ represents the measurement noise intrinsic to the acquisition process.

Inverse problems have wide applications in computer vision, medical imaging, optics, radar, and many other fields. The forward operator $\eH$ in (\ref{eqs:bases}) could represent various inverse problems,  from \egB an identity operator for image denoising, to convolution operators for image  deblurring, random sensing matrices for compressive sensing(CS), filtered subsampling operators for super-resolution (SR), (undersampled) Fourier transform for magnetic resonance imaging (MRI) and the (subsampled) Radon transform in computed tomography (CT). The inverse problems in (\ref{eqs:bases}) are often noisy and ill-posed since the operator $\eH$ has a non-trivial null space. Such under-determined systems are extremely difficult to solve and the solutions are very sensitive to the input data. The classical approach for solving them have traditionally been model-based \cite{boyd2011distributed,daubechies2004iterative}, which typically aim to regularize the solutions by constraining them to be consistent with prior knowledge about the signal and usually can only be solved iteratively.

More recently, due to the powerful representation learning and transformation ability, deep neural networks (DNN)  have emerged as a new paradigm for inverse problems. The community has already taken significant steps in this direction, with deep neural networks being successfully applied to a wide variety of inverse problems \cite{lucas2018using}. For example, \cite{burger2012image,xie2012image} use a fully connected feedfoward neural network for image denoisng and inpainting.  \cite{dong2015image,kulkarni2016reconnet} learn end-to-end mappings between $\ey_{\epsilon}$ and $\ex$ with vanilla convolutional neural networks (CNN). \cite{zhang2017beyond,ledig2017photo} further use CNNs with residual blocks \cite{he2016deep} and skip connections to improve the neural network performance. \cite{mao2016image,pathak2016context,jin2017deep} learn downsampling and upsampling feature maps with encoder-decoder CNNs. \cite{cui2014deep,zeng2015coupled} use autoencoders for learning new representations for $\ex$ and $\ey_{\epsilon}$ to solve the inverse problems. \cite{ulyanov2018deep}  use CNN as a prior and train with early stopping criteria to recover a single image from its observation. \cite{metzler2017learned} use CNN to perform denoising-based approximate message passing. \cite{sun2016deep,mardani2018neural} unfold the model-based optimizations with DNN. \cite{mousavi2015deep,bora2017compressed} use generative models for natural images to recover images from Gaussian measurements.

%ravishankar2019image,mccann2017convolutional

%More recently, due to the powerful representation learning and transformation ability, deep neural networks (DNN)  have emerged as a new paradigm for inverse problems. The community has already taken significant steps in this direction, with deep neural networks being successfully applied to a wide variety of inverse problems \cite{mccann2017convolutional,lucas2018using}. For example, \cite{burger2012image,xie2012image} use a fully connected feedfoward neural network for image denoisng and inpainting.  \cite{dong2015image,kulkarni2016reconnet} learn end-to-end mappings between $\ey_{\epsilon}$ and $\ex$ with vanilla convolutional neural networks (CNN). \cite{zhang2017beyond,ledig2017photo} further use CNNs with residual blocks \cite{he2016deep} and skip connections to improve the neural network performance. \cite{mao2016image,pathak2016context,jin2017deep} learn downsampling and upsampling feature maps with encoder-decoder CNNs. \cite{cui2014deep,zeng2015coupled} use autoencoders for learning new representations for $\ex$ and $\ey_{\epsilon}$ to solve the inverse problems. \cite{ulyanov2018deep}  use CNN as a prior and train with early stopping criteria to recover a single image from its observation. \cite{metzler2017learned} use CNN to perform denoising-based approximate message passing. \cite{mardani2018neural} unfold the model-based optimizations with DNN. \cite{bora2017compressed} use generative models for natural images to recover images from Gaussian measurements.

However, the DNN itself in the above deep learning-based approaches often lack the guarantees of  traditional physics-based methods as they are purely data-driven and learning-based. In addition, the designing of DNNs is usually complicated and has poor intuitive interpretation when they are decoupled from the inverse problem of interest. Furthermore, it is a commonly held belief in the inverse problems community that using the physics is preferable to relying solely on data \cite{schwab2018deepnull,mardani2017deep}. This raises a number of questions: is a purely data-driven neural network the best way to solve an inverse problem? Does physical information facilitate neural networks to find a better inverse problem solution? How should one best make use of the prior physical (acquisition) information? All of these above questions inspire us to think about whether the introduction of physical information in neural networks would be beneficial to the training of deep learning methods to better solve inverse problems.

We present \emph{Deep Decomposition learning Network} (DDN) as a way of using physics to reconstruct image from its measurement $\ey_{\epsilon}$ ($\epsilon\neq 0$).  Relying on the range-nullspace decomposition of data and the recent null-space learning \cite{mardani2017deep,schwab2018deepnull}, also known as affine projected network in \cite{sonderby2016amortised}, we propose to use two set of neural network layers to separately capture the residuals lying on the range of $\eHT$ and the nullspace of $\eH$. By incorporating the two learned residuals with the pseudo-inverse input, the proposed framework DDN is able to recover a plausible image and preserve data consistency with respect to the measurements $\ey_{\epsilon}$.

Decomposition learning can be used with various inverse imaging problems. In this paper, we follow a compressive sensing magnetic resonance fingerprinting (CS-MRF) reconstruction task \cite{davies2014compressed,ma2013magnetic}: reconstructing brain MR images sequence from highly undersampled and noisy Fourier space data. We show that the DDN outperforms recent related deep learning methods including deep residual learning \cite{jin2017deep} which is physics-free, deep nullspace learning  \cite{sonderby2016amortised,mardani2017deep,schwab2018deepnull} which neglected the noise component, and deep unrolled learning  \cite{gu2018integrating,gupta2018cnn,rick2017one} which retained the undesirable iterative nature of model-based systems. We also performed an ablation study on the natural image super-resolution (SR) task, and the results show that simultaneous learning of two residuals is outperforming the way that does not use physics or only learn one of the residuals. Finally, all the experiments demonstrate that decomposition learning can facilitate the generalization of the deep neural network for inverse problems.
%Our code is available at: \url{https://github.com/edongdongchen/DDN}.

\section{Background}\label{sec:background}

\subsection{Deep learning for the inverse problem}
Depending on whether the physical acquisition information with respect to $\eH$ is used during DNN training and testing, we divide the deep learning approaches into two categories: \emph{Physics-free} and \emph{Physics-engaged}.

\textbf{Physics-free.} The DNN aims to learn a direct mapping from $\ey_{\epsilon}$ (or its projection, e.g. $\eHT\ey_{\epsilon}$) to $\ex$ without exploiting the knowledge of $\eH$ at any point in the training or testing process (with the exception of the input). The general principle is that, given enough training data, we should be able to design a proper DNN to learn everything we need to know about $\eH$ to successfully estimate $\ex$ directly. The success of this approach is likely to depend on the complexity of the forward operator $\eH$. However, it has been observed to work well for numerous computer vision tasks, such as denoising and inpainting \cite{xie2012image}, superresoluion \cite{dong2015image} and deblurring \cite{xu2014deep,kupyn2018deblurgan}.  The DNN can be trained by a sole least squares loss  \cite{xie2012image,mao2016image,jin2017deep} or a combination of least squares loss and auxiliary losses such as adversarial loss \cite{ledig2017photo,kupyn2018deblurgan,quan2018compressed,shaham2019singan}.
% DELETED *** these are not solving inverse problems *** or trained without ground truth \cite{soltanayev2018training,zhussip2019training},
% DELETED *** DIP uses the physical model!!! *** or regard the DNN as a fixed prior \cite{ulyanov2018deep} to solve the inverse problem for a single image observation.
In general, this approach requires large quantities of training data because it is required to not only learn the geometry of the image space containing $\ex$, but also aspects of $\eH$. Hence, an interesting question is how hard (relatively) are each of these components to learn and how important is it to incorporate $\eH$ into the learning process. When the forward problem is too complex such that it can not be incorporated into the DNN model it will always be necessary to go Physics-free. Finally, since direct estimation using a DNN for solving inverse problems is essentially a form of regression, there is a potential generalization issue with such physic-free DNN approaches.
% physic-free deep learning approaches.

%zhang2019deep
\textbf{Physics-engaged.}
The most widely used strategy considering physics of $\eH$ in deep learning approaches is through a model-based approach, in which one or more pretrained DNNs are used within a more traditional iterative physics-engaged model-based framework such as \cite{boyd2011distributed,daubechies2004iterative}. As mentioned before, the inverse problem (\ref{eqs:bases}) typically boils down to solving an optimisation problem broadly of the following form:
    \begin{equation}\label{eqs:pnp_regularization}%\label{loss:inverse}
     \arg\min_{\ex}f(\ex) + \lambda\phi(\ex),
    \end{equation}
where the first term $f(\ex)$ aims to enforce data fidelity, \egB $f$ could be the MSE between $\eH\ex$ and $\ey_{\epsilon}$, while the regularizer  $\phi$ allows us to insert knowledge onto the solution $\ex$, and $\lambda\in \eR^+$ controls the strength of the regularization. Typically there is no closed-form solution to (\ref{eqs:pnp_regularization}) and it usually needs to be solved iteratively. This has led to the following proposed uses for pretrained DNNs: (i) use DNN to replace the proximal operator associated with $\phi(\ex)$ in a proximal gradient algorithm \cite{gu2018integrating,gupta2018cnn,rick2017one}, (ii) use DNN to replace the gradient $\nabla\phi$ in an unrolled optimization method
\cite{chen2016trainable,sun2016deep,mardani2018neural}, (iii) directly replace the regularizer $\phi$ with DNN \cite{lunz2018adversarial,romano2017little}, (iv) use DNN as a generative model to generate $\ex$ from a latent code  that needs to be estimated \cite{bora2017compressed,shah2018solving}. These iterative methods are Physics-engaged, as they actually use the regularizer along with the forward model and observation by minimizing the disparity between the oracle and its reconstruction.

As an exception to the above physics-engaged deep learning approaches, there have been some recent studies aimed at explicitly using $\eH$-related information during the DNN training process in an end-to-end manner. For example, \cite{schwab2018deepnull,mardani2017deep} explicitly learn the nullspace component of $\ex$ with respect to $\eH$. However, this separate nullspace learning does not deal with the presence of noise in the input nor with situations where no nullspace exists. Another interesting direction presented in \cite{gilton2019neumann} considers a Neumann series expansion of linear operators to approximate the inverse mapping of (\ref{eqs:pnp_regularization}). However, this requires the network to precondition and store the results of each iteration in order to accumulate the approximate Neumann series to solve the inverse problem.

In this paper, inspired by the nullspace method of \cite{mardani2017deep,schwab2018deepnull,sonderby2016amortised}, we explore the possibility of a more flexible end-to-end neural network structure that is capable of exploiting both the range and null space structures of the inverse problem. Before discussing the proposed method, let us briefly recall the Range-Nullspace decomposition of data.

\subsection{Range-Nullspace ($\R$-$\N$) Decomposition}
Given a linear forward operator $\eH\in\eR^{d\times D}$ and its right pseudo inverse $\eHT\in\eR^{D\times d}$, which satisfies $\eH\eHT=\I_d$, it holds that  $\eR^D = \R(\eHT)  \oplus \N(\eH)$, which  implies that for any sample $\forall \ex\in \eR^D$ there exists two unique elements $\ex^{+}\in \R(\eHT)$ and $\ex^{\perp}\in \N(\eH)$ such that $\ex=\ex^{+}+\ex^{\perp}$. Therefore we define the following range-nullspace ($\R$-$\N$) decomposition,
\begin{definition} \textbf{$\mathcal{R}$-$\mathcal{N}$ Decomposition:}
Let $\mathcal{P}_r\triangleq \eHT\eH$ be the operator that projects the sample $\ex$ from sample domain to the range of $\eHT$, and denote by $\mathcal{P}_n\triangleq(\I_D-\eHT\eH)$ the operator that projects $\ex$ to the null space of $\eH$. Then $\forall\ex\in \eR^D$, there exists the unique decomposition:
\begin{equation}\label{eqs:ns-decom}
    \ex = \PP_r(\ex) + \PP_n(\ex),
\end{equation}
where we will call $\PP_r(\ex)$ and $\PP_n(\ex)$ the \emph{$r$-component} and \emph{$n$-component} of $\ex$, respectively.
\end{definition}

\begin{remark}
In this paper we will only focus on the above decomposition. However, we comment that in principle the pseudo-inverse, $\eHT$, could be replaced by any general right inverse of $\eH$ in the above decomposition which might provide added flexibility in certain inverse problems.
\end{remark}

% An illustration of $\R$-$\N$ Decomposition is shown in Figure \ref{fig:example_rn_dec}.
Thus, the task of solving an inverse problem is to find these two components $\PP_r(\ex)$ and $\PP_n(\ex)$ based on the observed data, $\ey_{\epsilon}$.  The simple linear estimator to solve this problem is to use the approximation:
\begin{equation}\label{eqs:ns-decom-bp}
\ex^* = \eHT\ey_{\epsilon}.
\end{equation}
This estimator enjoys global and exact data-consistency, \ieB $\eH\ex^*\equiv\ey_{\epsilon}$, which is an important consideration when solving inverse problems \cite{mardani2017deep}. However, comparing  (\ref{eqs:ns-decom-bp}) with (\ref{eqs:ns-decom}) we can see that this is achieved by simply setting the nullspace component to zero: $\PP_n(\eHT\ey_{\epsilon})=0$. In general this provides a poor solution for ill-posed problems. Thus it is necessary to further estimate the missing component $\PP_n(\ex)$. Such an estimator is necessarily nonlinear.

\textbf{Nullspace learning.} Recently, \cite{mardani2017deep,schwab2018deepnull,sonderby2016amortised} explored the use of a neural network $\G$ to feed a refined backprojection $\G(\eHT\ey_{\epsilon})$ to the null-space projection operator $\PP_n$, then the reconstruction in (\ref{eqs:ns-decom-bp}) is reformulated as
\begin{equation}\label{eqs:nsn_rec}
\ex^* = \eHT\ey_{\epsilon}+\PP_n(\G(\eHT\ey_{\epsilon})),
\end{equation}
where the network $\G$ is suggested to be trained by minimizing the error between $\ex$ and  $\ex^*$. Note the solution (\ref{eqs:nsn_rec}) enjoys global data consistency, \ieB $\eH\ex^*\equiv\ey_{\epsilon}$. However, the solution (\ref{eqs:nsn_rec}) unfortunately, only works for the noise-free situation, and does not allow any denoising in the range $\mathcal{R}(\eHT)$.
%which does not denoise in the range  $\mathcal{R}(\eHT)$ but only learns the $n$-component in the null space  $\mathcal{N}(\eH)$. This makes it not suitable for inverse problems when the measurement data are noisy  ($\epsilon>0$).
Indeed, (\ref{eqs:nsn_rec}) can only denoise in the nullspace and the denoising ability is therefore worst-case bounded by $\|\epsilon\|/\|\eH\|$ since $\|\eH(\ex-\ex^*)\|=\|\epsilon\|$. The noise may further limit the ability to predict the null space component from the noisy measurements. Although it is reminiscent of decoupling the neural network denoiser from the inverse problem, it does not benefit from this since the training needs to be tailored to the task \cite{rick2017one}, which will be confirmed  in our experiments.
%Furthermore, this is highlighted by the fact that the nullspace network fails to solve the inverse problems where the nullspace $\N(\eH)$ does not exist, \egB the (trivial inverse) denoising problem in which $\eH=\I$.

\section{Deep Decomposition Learning}

\begin{figure*}[t]
     \centering
          \begin{subfigure}[b]{0.34\textwidth}
         \centering
         \includegraphics[width=\textwidth]{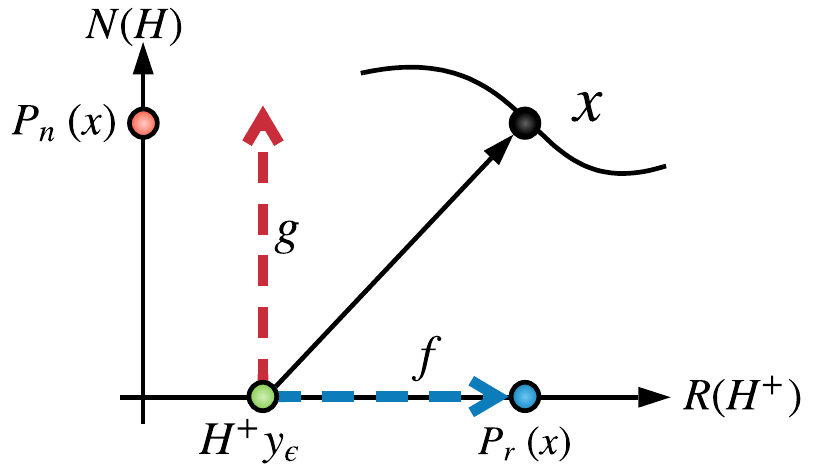}
         \caption{$\R$-$\N$ Decomposition}
         \label{fig:motivation_a}
     \end{subfigure}
     \begin{subfigure}[b]{0.27\textwidth}
         \centering
        \includegraphics[width=\textwidth]{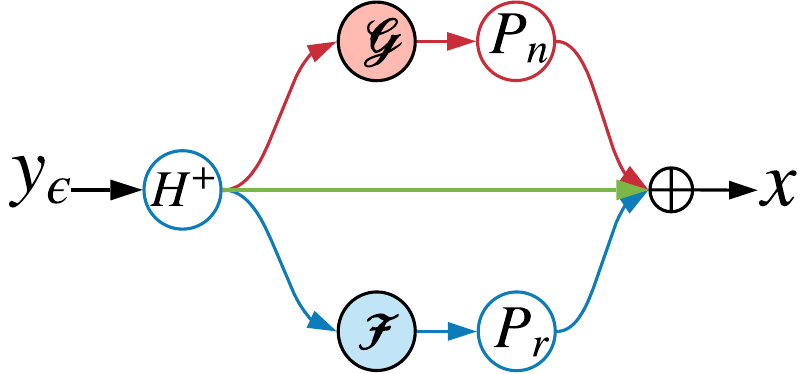}
         \caption{Independent}
         \label{fig:motivation_b}
     \end{subfigure}
    %  \hfill
    % \hspace{5pt}
     \begin{subfigure}[b]{0.37\textwidth}
         \centering
        \includegraphics[width=\textwidth]{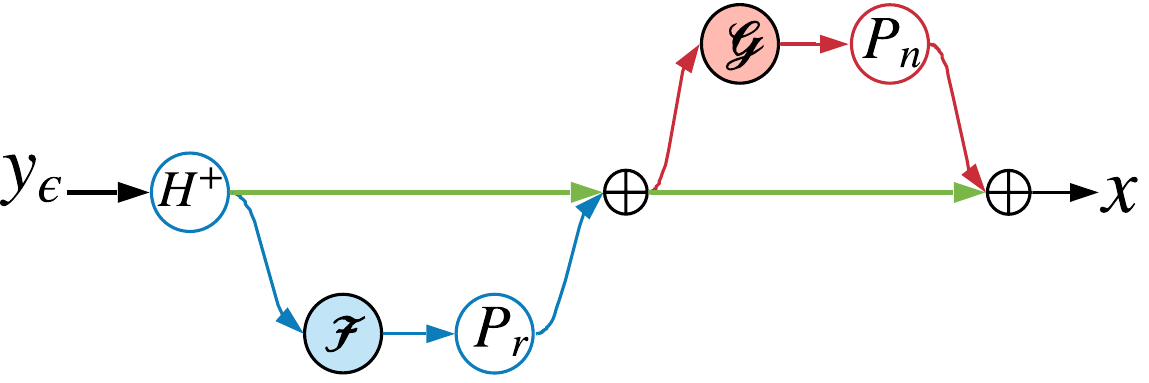}
         \caption{Cascade}
         \label{fig:motivation_c}
     \end{subfigure}
        \caption{Decomposition learning (a) and its two corresponding network architectures in (b) and (c). Decomposition learning trains neural network to recovery the range component $\PP_r(\ex)=\eHT\eH\ex$ and nullspace component $\PP_n(\ex)=\ex-\PP_r(\ex)$ of oracle image $\ex$ from its  coarse reconstruction $\eHT\ey_{\epsilon}$. The residuals between $\eHT\ey_{\epsilon}$ and $\PP_r(\ex)$ and $\PP_n(\ex)$ are recovered by $f=\PP_r(\mathcal{F})$ and $g=\PP_n(\mathcal{G})$ respectively,
        %The residuals between $\PP_r(\ex)$ and $\PP_n(\ex)$ and $\eHT\ey_{\epsilon}$ are recovered by $f=\PP_r(\mathcal{F})$ and $g=\PP_n(\mathcal{G})$,
        where $\mathcal{F}$ and $\mathcal{G}$ are two sets of neural layers need to train. Different from direct learn the residual between $\ex$ and $\eHT\ey_{\epsilon}$ without using physics, $\ex$ is instead reconstructed by $\eHT\ey_{\epsilon} + f + g$ in a decomposition learning mechanism. This method enables physics to engage in neural network training and testing and generate more physics plausible reconstruction.}
        \label{fig:motivation}
% \vspace{-5pt}
\end{figure*}

Inspired by nullspace learning \cite{sonderby2016amortised,schwab2018deepnull,mardani2017deep} we aim to remove the range space denoising deficiency while still exploiting the nullspace property. Let us consider the case $\epsilon\neq 0$ in (\ref{eqs:bases}) which is more practical. By the $\R$-$\N$ decomposition (\ref{eqs:ns-decom}), it holds that $\ex$ can be exactly recovered by,
\begin{equation}\label{eqs:dec-01}
\ex = \eHT\ey_{\epsilon} - \eHT\epsilon + \PP_n(\ex).
\end{equation}
However, as mentioned before, in the scenario of the inverse problem, both the oracle image $\ex$ and the noise term $\epsilon$ in (\ref{eqs:dec-01}) are still unknown and need to be recovered.

We address this problem by using two neural networks and introducing the decomposition learning framework (illustrated in Fig.\ref{fig:motivation_a}). Instead of hoping a single neural network will directly fit a desired underlying mapping between $\eHT\ey_{\epsilon}$ and $\ex=\PP_{r}(\ex)+\PP_{n}(\ex)$, we explicitly let two networks, denoted by $\F$ and $\G$ fit the two mappings from $\eHT\ey_{\epsilon}$ to the residual $r$-component $\PP_{r}(\ex)-\eHT\ey_{\epsilon}$ and the $n$-component $\PP_{n}(\ex)$, respectively. In particular, the output of $\F$ should be bounded by the magnitude of noise $\epsilon$, while $\G$ should be a smooth, \ieB a Lipschitz continuous neural network, since $\G$ is essentially a nullspace network, which does not need to be strongly bounded but should be regularized in order to get reasonable generalization. Therefore, the oracle $\ex$ is decomposed as the sum of a linear component $\eHT\ey_{\epsilon}$  (the input), a bounded residual component $\PP_r\circ\F\in \R(\eHT)$ and a smooth $n$-component $\PP_n\circ\G \in \N(\eH)$.
%The decomposition learning is illustrated in Fig.\ref{fig:motivation_a}.

We consider two versions of DDN estimators. First, we define an independent connection architecture (Fig.\ref{fig:motivation_b}) estimator $\A_i$ using the $\mathcal{R}$-$\mathcal{N}$ decomposition,
\begin{equation}\label{eqs:dec_rec}
\A_i(\ey_{\epsilon}) \triangleq  \eHT\ey_{\epsilon} + \PP_r(\F(\eHT\ey_{\epsilon}))+ \PP_n(\G(\eHT\ey_{\epsilon})).
\end{equation}
where there are no interactions between $\F$ and $\G$.

An alternative (but essentially equivalent) mapping $\A_c$ from $\ey_{\epsilon}$ to $\ex$ which is related to (\ref{eqs:dec_rec}) uses a cascade of networks (Fig.\ref{fig:motivation_c}), \ieB first denoising with $\F$, then feeding the denoised $\eHT\ey_{\epsilon} + \PP_r(\F(\eHT\ey_{\epsilon}))$ into $\G$ such that:
\begin{equation}\label{eqs:dec_rec_cascade}
% \begin{split}
    \A_c(\ey_{\epsilon})  \triangleq \eHT\ey_{\epsilon} + \PP_r(\F(\eHT\ey_{\epsilon})) + \PP_n(\G  ( \eHT\ey_{\epsilon}+ \PP_r(\F(\eHT\ey_{\epsilon})))).
% \end{split}
\end{equation}
Intuitively (\ref{eqs:dec_rec_cascade}) is preferable since it is no more complex than the independent network topology and provides the nullspace network $\G$ with a range denoised input, thereby reducing the learning burden needed to be done by $\G$. Our experiments also verified the cascade connections typically perform better than the independent one. Note that the independent model is by its nature a shallower but wider network than the cascade model which by construction is deeper, but given the decomposition both networks have essentially the same complexity.

%Fig.\ref{fig:motivation_b}-\ref{fig:motivation_b} gives an illustration of the above deep decomposition learning, as well as the corresponding independent architecture and the cascade architecture for $\F$ and $\G$  in (\ref{eqs:dec_rec}) and (\ref{eqs:dec_rec_cascade}), respectively.

The $\mathcal{R}$-$\mathcal{N}$ decomposition learning gives the outputs of the two networks $\F$ and $\G$ clear interpretability: the output from $\F$ estimates the noise component, while the output from $\G$  represents the component that is inferred from the implicit image model as it was not measured at all by $\ey_{\epsilon}$. Second, the DDN estimator $\A(\ey_{\epsilon})$ defined in (\ref{eqs:dec_rec}) and (\ref{eqs:dec_rec_cascade}) offers the ability to denoise both the $r$-component  $\PP_{r}(\ex)$ and the $n$-component $\PP_{n}(\ex)$, and if $\|\eH\F\|\leq\|\epsilon\|$, the solution enjoys a relaxed notion of data-consistency (in the spirit of the discrepancy principle popular in inverse problems), which convergences to exact data consistency when $\|\eH\F-\epsilon\|\rightarrow 0$, that makes the deep learning solutions more physically plausible. Finally, in the cascade architecture (Fig. \ref{fig:motivation_c}) we calculate $\PP_r(\F)$ first. If we had cascaded $\PP_n (\G(\eHT\ey_{\epsilon})) + \eHT\ey_{\epsilon}$ first we would have lost the $\mathcal{R}$-$\mathcal{N}$ decomposition as the input to $\F$ would live in the full space. The input to $\G$~would also have range noise, as in the independent architecture, thereby increasing the training burden on $\G$ network.

%Note if we first applied $\PP_n\circ \G$ we would only retain the null space component and may have irretrievably lost information from the measurements $\ey_{\epsilon}$.
%
%if we first calculated $PP_n (\G(\eHT\ey_{\epsilon})) + \eHT\ey_{\epsilon}$ we would have lost the $\mathcal{R}-\mathcal{N}$ decomposition as the input to F would live in the full space.
%
%The input to $\G$ would also have range noise, as in the independent architecture, thereby increasing the training burden on this network.

% This would not be equivalent. Alternatively, and the case we believe the reviewer is alluding to, if we first calculated P_n (G(H^\dagger y)) + H^\dagger y we would have lost the R-N decomposition as the input to F would live in the full space. The input to G would also have range noise, as in the independent architecture, thereby increasing the training burden on this network.

%The interpretability is illustrated in two ways: first, by incorporating the physics into the network, the solutions enjoy data consistency that makes the deep learning solutions more physically plausible. Second, the $\mathcal{R}$-$\mathcal{N}$ decomposition gives the outputs of the two networks $\F$ a $\G$ a clear meaning: the output from $\F$ estimates the noise component, while the output from $\G$ represents the component that is inferred from the implicit image model as it was not measured at all by $\ey_{\epsilon}$.

\subsection{Training strategy}
Let $\eX = \{(\ey_{\epsilon}^{(i)}, \ex^{(i)})\}_{i=1}^N$ denote a training set  of $N$ samples, where $\ex^{(i)}$ and $\ey_{\epsilon}^{(i)}$ are the clean oracle signal and its noisy measurement given by (\ref{eqs:bases}). Denote by $\ell(x, y)$ the loss function, which measures the discrepancy between $x$ and $y$. In this paper we used $\ell_2$ to compute $\ell$. Given an estimator $\A$, we jointly training $\F$ and $\G$  by solving a single optimization program,
\begin{equation}\label{eqs:ddn_join_loss}
    \min\limits_{\F,\G}\ell_{\text{emp}}(\A) + \lambda_1\phi_1(\F) + \lambda_2\phi_2(\G),
\end{equation}
where the first term $\ell_{\text{emp}}(\A)\triangleq \frac{1}{N}\sum_{\ey_{\epsilon}^{(i)}, \ex^{(i)}\in\eX}\ell(\A(\ey_{\epsilon}^{(i)}), \ex^{(i)})$ is $\A$'s empirical loss associated with the training set $\eX$ and serves the data-fidelity, and $\A$ take the form either in (\ref{eqs:dec_rec}) or in (\ref{eqs:dec_rec_cascade}). In this paper, we set $\phi_1(\F) = \sum_{i=1}^N\ell(\eH\F(\eHT\ey_{\epsilon}^{(i)}), \epsilon^{(i)})$ in order to encourage the data discrepancy term to be small. We then set $\phi_2$ as the weight decay term to control the Lipschitz of the network \cite{schwab2018deepnull} and to encourage good generalization \cite{kawaguchi2017generalization}. It worth noting we can flexibly define different $\phi_1$ and $\phi_2$ for $\F$ and $\G$ to impose the desired boundedness and smoothness conditions, such that to tune the networks to their specific tasks.

Due to the decomposition $\ex = \PP_r(\ex) + \PP_n(\ex)$ in the independent architecture there are no interactions between the respective targets of $\F$ and $\G$, therefore one can decouple (\ref{eqs:ddn_join_loss}) into two independent sub-optimizations to train $\F$ and $\G$ separately, \ieB $\min_{\F}(\PP_r(\F), \PP_r(\eX))+\lambda_1\phi_1$ and $\min_{\G}(\PP_n(\G), \PP_n(\eX))+\lambda_2\phi_2$. While joint training and decoupled training are theoretically equivalent, in practice, the decoupled training enjoys more intuitive interpretability, and it is easier to control $\F$ and $\G$ each to achieve better convergence results. However, the joint training is slightly more efficient than the decoupled  because the networks can be trained simultaneously. Accordingly we use the cascade architecture as our DDN in our experiments and jointly train it using (\ref{eqs:ddn_join_loss}).

\subsection{The relationship to other work}
In the noise-free case ($\epsilon=0$), the decomposition learning (\ref{eqs:dec-01}) reduces naturally to vanilla nullspace learning (\ref{eqs:nsn_rec}). Thus a DDN can be regarded as a generalized nullspace network. While in the noisy case, one might be tempted to consider adopting a separate generic denoiser to preprocess the measurements $\ey_{\epsilon}$. However, such denoisers are typically built in a manner decoupled from the inverse problem of interest, therefore to train a nullspace network with such denoised measurements could amplify the reconstruction error, causing more inconsistent results \cite{rick2017one}. In contrast, the denoising process in the DDN is not decoupled from the inverse problem but integrated into a unified learning model. The experiments show this not only improves the quality of the results but in addition, helps the model's generalization.  Note DDN is different from the recent heuristic decomposition network \cite{pan2018learning} which broadly separates the image into structure part and detail part. In contrast, we identify the natural $\mathcal{R}$-$\mathcal{N}$ decomposition induced by $\eH$ that also allows us to select appropriate loss functions for the different components.

Our decomposition learning can also be regarded as a special gated neural network. To be specific, if we rewrite (\ref{eqs:dec_rec}) as
\begin{equation}
    \A = \eT\F(\ez) + (\I-\eT)\G(\ez) + \ez,
\end{equation}
where $\ez=\eHT\ey_{\epsilon}$, $\eT=\eHT\eH$, it can be seen the output of $\F$ and $\G$ are gated in terms of $\eT$ and $\I-\eT$. The importance of the two components is determined by the physics in terms of $\eHT\eH$. This is different from previous gated networks such as
\cite{hochreiter1997long,srivastava2015training} in which the model is gated by some bounded numerical function such as a sigmoid or hardlim which are not typically related to the physics of the forward model or its inverse. Our method can also be regarded as a generalized residual learning \cite{he2016deep}, \ieB we decompose the residual $\A-\ez$ into two components in $\R(\eHT)$ and $\N(\eH)$ with more explicit interpretability. In particular, in the absence of the nullspace $\N(\eH)$ or in the case $\eH=\I$ such that $\PP_r = \I$ and $\PP_n=0$, \ieB there is no nullspace learning, $\G$ will be irrelevant and only $\F$ will be learnable, and the decomposition learning will be reduced to a non-gated neural network and equivalent to the standard residual learning.

% (in Pan simple L2 losses are used).
%
%There is indeed similarity between Pan’s dualCNNs and our networks. However, there are also key differences. Pan et al propose heuristic decompositions for different tasks and broadly separate the image into structure S and detail D. In contrast, we identify the natural R-N decomposition induced by H that also allows us to select appropriate loss functions for the different components (in Pan simple L2 losses are used).
%
%In general, the S-D decomposition is not comparable with the R-N decomposition proposed here. The example where the two methods are closest is SR where Pan et al define S as a Gaussian filter of X and D=X-S as the difference, which crudely approximates the range and null spaces. We will highlight these points in the final version and show our two SR reconstructed components separately (as suggested by Reviewer 2) for comparison.

\begin{remark}
A key aspect of the DDN framework is access to the projection operators $\PP_n$ and $\PP_r$. The complexity will depend on how easy it is to approximate $\eHT$, or another left inverse.
%If the inverse problem is relatively small, then $\eHT$ can be calculated directly or approximated using truncated SVD. However, in larger scale problems this will not be feasible and alternative approximations must be sought.
%For many computer vision problems, such as the ones considered in this paper - compressed sensing and super-resolution - the pseudo-inverse is readily available or easy to approximate.
While many computer vision problems admit an easily computable inverse, such as: deblurring, inpainting and the ones considered in this paper - compressed sensing and super-resolution, for more general inverse problems this approximation is more challenging and may even be unstable. However, it is possible to approximate $\ez=\eHT\ey$ by a regularized linear operator, e.g. $\arg\min_{\ez}  \|\ey-\eH\ez\|^2 + \lambda \|\ez\|^2$. Here we can leverage the wealth of literature in inverse problems dedicated to efficiently computing this, e.g. using preconditioned conjugate gradient solvers, etc. \emph{Unrolled} physics-engaged deep learning solutions \cite{gu2018integrating,gupta2018cnn,rick2017one} also solve a similar optimization problem but with an integrated complex DNN regularizer, and therefore resort to slower proximal first order methods. In contrast, the DDN allows us to separate the DNN mappings from this optimization. For such challenging problems we doubt whether a similar complexity physics-free solution would be competitive.

\end{remark}

\section{Experiments}
We designed our experiments to address the following questions:
\begin{itemize}
    \item \emph{Does the proposed deep decomposition learning help the neural network produce superior results on different inverse problem tasks? Does the DDN enjoy better generalization?}
    \item \emph{Which training strategy, e.g. jointly or decoupled, is best for training the DDN? Which connection type,
\egB the independent or cascade one, is better?}
    \item \emph{How important is each component $\PP_r(\F)$ and $\PP_n(\G)$ to DDN?}
\end{itemize}
To do that, we validate the effectiveness of our proposed decomposition learning on the inverse problems of compressive sensing MR fingerprinting (CS-MRF) \cite{ma2013magnetic,davies2014compressed} reconstruction and natural images super-resolution. While our focus is mostly on CS-MRF, the SR experiments are included to shed some light on the ablation analysis.

\begin{table}[t]
     \caption{Four DNN solvers $\A$ that use different mechanisms for using physics to solve inverse problems. Note the residual learning is physic-free and $\text{Prox}_{\eH}$ is the proximal operator in proximal gradient descent optimization.}%and the physics is not direct used neural network training. $\I$ is the identity matrix, $\G$ and $\F$ are two set of neural layers.
        \centering
        \fontsize{8}{8}\selectfont %\fontsize{6.5}{5.3}{7.0}{6.0} {fontsize}{vspace}
        \begin{threeparttable}
            \begin{tabular}{l |c |c |c |c}
            \toprule
            Mechanisim  &Residual learning	&Unrolled learning	&Nullspace learning &Decomposition learning\\ \midrule
            Formulation&$\A=\I+\G\circ\F$	&$\text{Prox}_{\eH}\leftarrow \G\circ\F$ &$\A=\I+\PP_n(\G\circ\F)$	&$\A=\I+\PP_r(\F)+\PP_n(\G)$\\
            \bottomrule
            \end{tabular}
        \end{threeparttable}
\label{table:mechanism}
% \vspace{-5pt}
\end{table}

\subsection{Implementation}
Our goal here is not to explore the potential designs of the neural networks, $\F$ and $\G$, but the usage of physics in the neural network. Therefore, we use a simple four layers CNN as $\F$ and directly apply UNet \cite{ronneberger2015u}, which is commonly used in the field of image reconstruction \cite{jin2017deep,gupta2018cnn,ulyanov2018deep}, as $\G$. In particular, all the layers in $\F$ are with $3\times 3$ kernels where the first three layers undergo ReLU activation and the 2nd and 3rd layers were followed by batch normalization and ReLU activation, and a fixed number of $c$, 64, 64, $c$ feature maps for each layer where $c$ is the number of input channels. We compare the proposed \emph{decomposition learning} (DDN) with \emph{residual learning} (ResUnet, \cite{jin2017deep}), \emph{unrolled learning} (NPGD, \cite{mardani2018neural}) and \emph{nullspace learning} (NSN, \cite{meinhardt2017learning,schwab2018deepnull}). The above different mechanisms are summarized in Table \ref{table:mechanism}. We build NPGD with a fixed 3 recurrent steps \cite{mardani2018neural}. It is important to note that we keep the base neural network topology and the number of parameters the same for both our DDN and the competitors.

We quantify the image reconstruction quality using Normalized MSE (NMSE), Peak Signal to Noise Ratio (PSNR), \ieB $\text{PSNR}=10\log_{10}(255/\text{MSE})^2$, and generalization error (GE)-the difference between the expected loss and the training loss, is evaluated by $\text{GE}(\A) = |\ell(\A,\X_{\text{test}})-\ell(\A, \X_{\text{train}})|$ where $\ell(\A, \X)$ denotes the loss evaluated over the data $\X$. We used ADAM to optimize the DDN and tuned the $\lambda_1$ and $\lambda_2$ for specific inverse problems. All networks used $\eHT\ey_{\epsilon}$ as the input, were implemented in PyTorch, and trained on NVIDIA 2080Ti GPUs.

\subsection{CS-MRF reconstruction}

Magnetic Resonance Fingerprinting (MRF) \cite{ma2013magnetic} recently emerged to accelerate the acquisition of tissues’ quantitative MR characteristics include $T_1$, $T_2$, off-resonance frequency, etc. The compressive sensing MRF (CS-MRF) \cite{davies2014compressed} is typically involves two goals, reconstructing a reasonable MR fingerprint (multi-coil sensitivities, \ieB MR images sequence) $\ex$ from its undersampled k-space measurement $\ey_{\epsilon}$, and at the same time using pattern recognition techniques such as dictionary matching \cite{ma2013magnetic} or neural networks \cite{fang2019deep,chen2019bMIDL,mo2019geometry}, to query the corresponding quantitative map $\emm$, such that achieve quantification of MRI. Accordingly, the data acquisition of CS-MRF can be boiled down to $\ey_{\epsilon} =\eH(\ex) + \epsilon$ where the forward operator $\eH$ consists a Fourier transformation $\F$ followed by a per-frame and time-varying subsampling operator $\eS$, \ieB $\eH=\eS\circ\F$. For the simplicity, we use the Cartesian sampling as example and now $\eHT =  \F^{-1}\circ \eS^{\top}$.

Note that $\ex$ can be generated by the Bloch equation using its tissue property/parameters map $\emm$, thus predicting $\emm$ from $\ex$ is another inverse problem and is nonlinear. Recent studies show that this nonlinear inverse problem $\ex\rightarrow \emm$ can be effectively solved by neural networks \cite{fang2019deep,chen2019bMIDL,mo2019geometry}. Here we mainly focus on the inverse problem $\ey_{\epsilon}\rightarrow\ex$, \ieB recover $\ex$ from its noisy measurement $\ey_{\epsilon}$. We then use the UNet-based network \cite{fang2019deep} that pre-trained on the clean oracle MRF images to predict the tissue property map $\emm$ for a given MRF image $\ex$.

\subsubsection{Dataset.} We use the simulated human brain MRF dataset in \cite{chen2019bMIDL}. The ground truth parametric tissue property maps are collected from 8 volunteers (2k slices) using MAGIC \cite{magic2015} quantitative MRI protocol with Cartesian sampling. These parametric maps are then used to simulate MRF acquisition using the  Fast imaging with Steady State Precession (FISP) \cite{jiang2015mr} protocol and Cartesian sampling. In particular, we set the dimension reduced 10 channel MRF image $\ex$ with scale $D=200\times 200$, the compression ratio is $d/D=1/40$, and the noise level $\sigma_{\epsilon}=0.01$.
In order to evaluate the generalization ability of different methods under the extremely case that \emph{scarce} labelled anatomical quantitative MRI datasets that are available for training,
%In order to evaluate the generalization ability of different methods,
small numbers of samples from the first 7 patients are randomly picked for training, \egB $N=100, 50, 10$ in our experiments and 20 slices from the last volunteer are picked for the test.

\subsubsection{Comparison results}

\begin{table}[t]
     \caption{Comparison results on CS-MRF reconstruction task.}
        \centering
        \fontsize{8}{12}\selectfont %\fontsize{6.5}{5.3}{7.0}{6.0} {fontsize}{vspace}
        \begin{threeparttable}
            \begin{tabular}{l |c| c | c| c | c |c| c}
            \toprule
              \multirow{2}*{}&IT &\multicolumn{3}{c|}{$X$ (NMSE$\times 10^{-2}$/GE)}	&\multicolumn{3}{c}{$T_1$ (PSNR dB)}\\  \cline{3-8}
               & (sec)&$N=100$ & $N=50$ & $N=10$	& $N=100$       & $N=50$ & $N=10$\\ \midrule
            ResUnet\cite{jin2017deep}& \textcolor{blue}{0.005}  &$0.62/0.027$    & $0.743/0.069$   & $1.119/0.164$     &$17.10\pm2.64$  &$17.80\pm1.93$   &$15.18\pm1.49$\\
            NPGD\cite{mardani2018neural}&  0.15   &$0.58/0.024$  & $0.980/0.118$     & $1.689/0.255$	        &$18.05\pm2.02$  &$15.49\pm1.47$   &$13.67\pm1.31$\\
            NSN\cite{schwab2018deepnull}& 0.04    &$0.57/0.031$  & $0.740/0.068$   & $1.119/0.159$	        &$18.49\pm2.39$  &$18.03\pm1.65$   &$15.25\pm 1.53$\\
            DDN & 0.06    &$\textcolor{blue}{0.51}/\textcolor{blue}{0.019}$ & $\textcolor{blue}{0.711}/\textcolor{blue}{0.062}$   & $\textcolor{blue}{1.056}/\textcolor{blue}{0.139}$    	& \textcolor{blue}{$18.92\pm2.04$}  & $\textcolor{blue}{18.06\pm1.54}$ &\textcolor{blue}{$15.90\pm1.49$}\\
            \bottomrule
            \end{tabular}
        \end{threeparttable}

\label{table:tab_csmrf}
% \vspace{-5pt}
\end{table}

\begin{figure*}[t]%\label{fig:debluring_example}
\centering
\begin{subfigure}{0.49\linewidth}
\centering
\rightline{\includegraphics[width=0.25\linewidth]{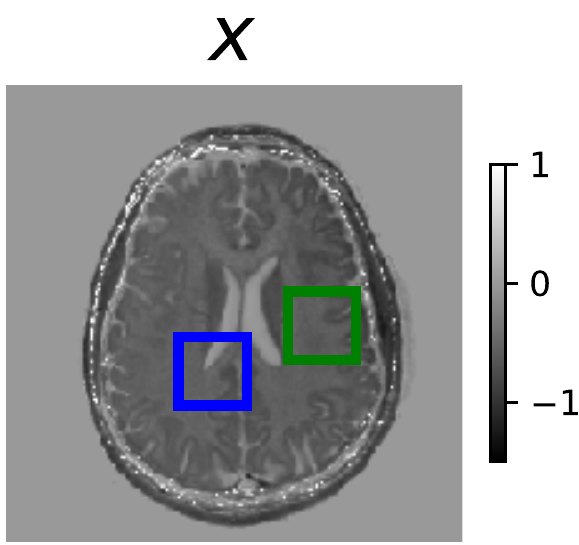}}
\rightline{\includegraphics[width=0.25\linewidth]{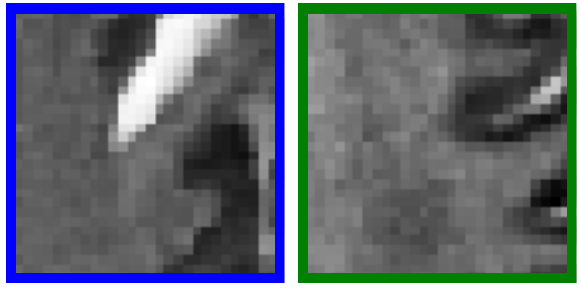}}
%\leftline{
%\includegraphics[width=0.23\linewidth]{./figs/x_gt.pdf}
%\includegraphics[width=0.24\linewidth]{./figs/x_gt_crop.pdf}
%}
\includegraphics[width=1\linewidth]{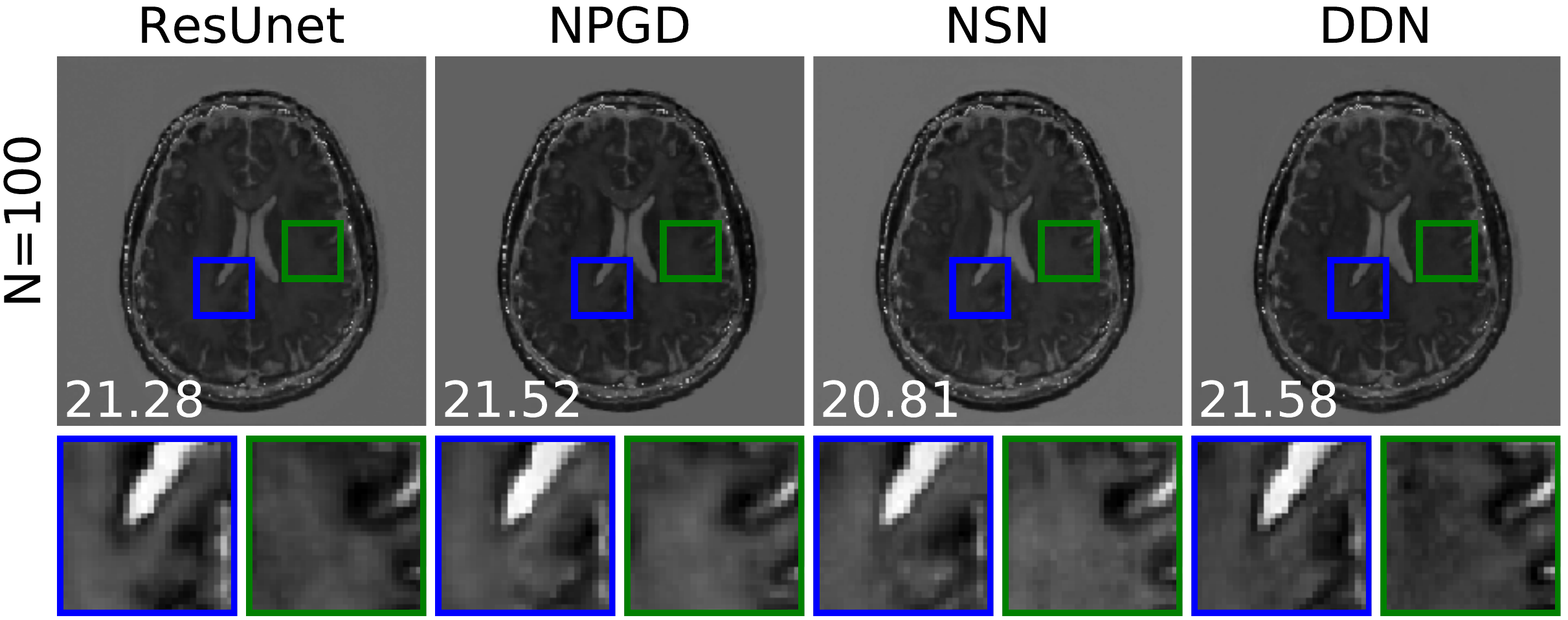}
\includegraphics[width=1\textwidth]{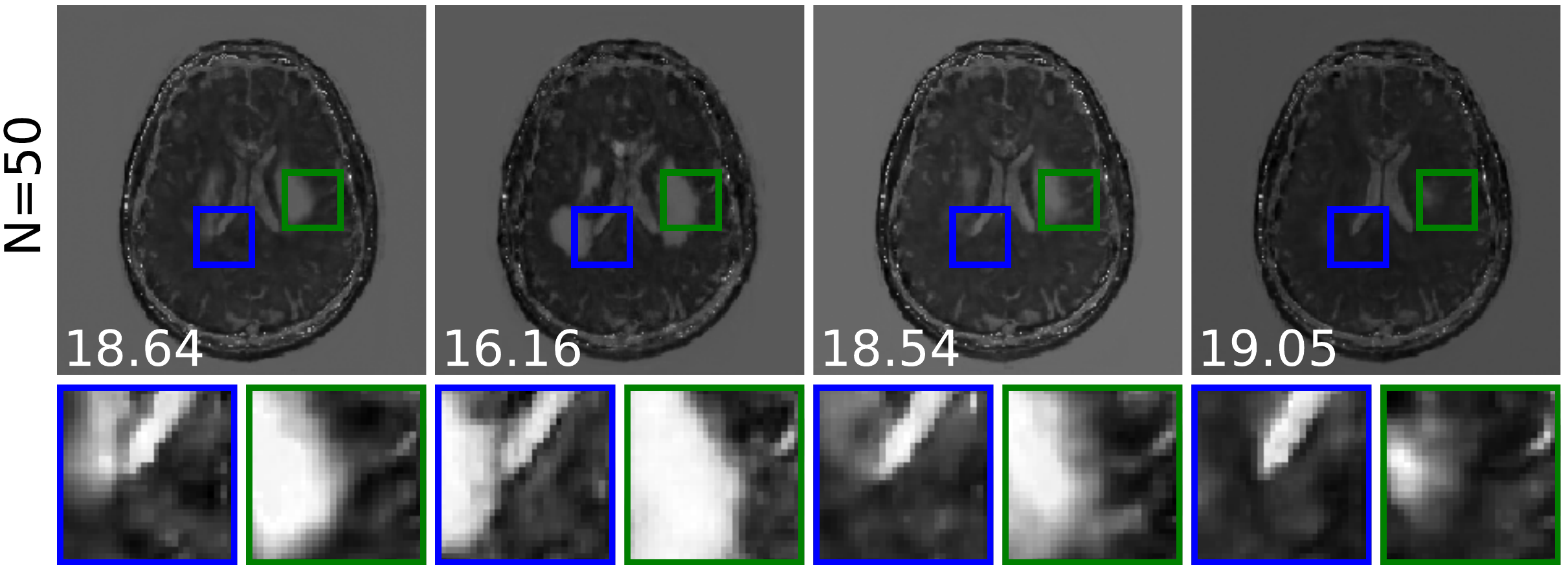}
\includegraphics[width=1\textwidth]{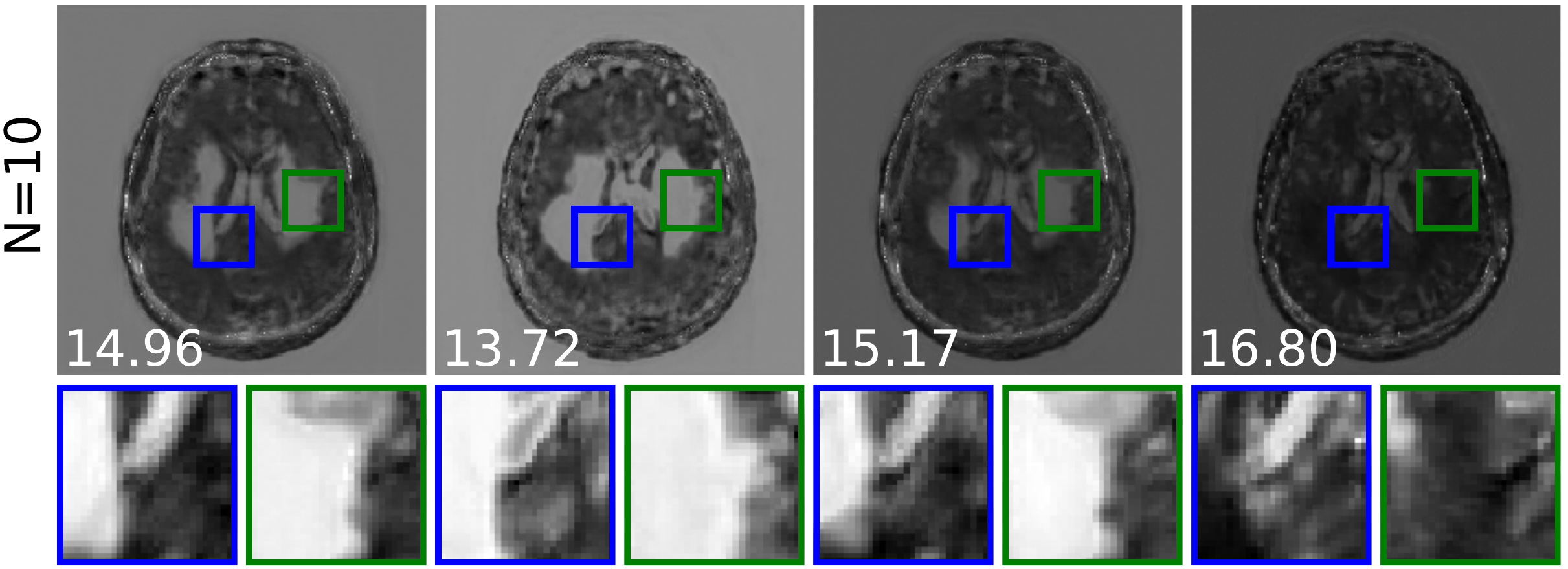}
  \caption{reconstruction of $\ex$.}
  \label{fig:reconstruction_x}
\end{subfigure}
\begin{subfigure}{0.49\linewidth}
\centering
%\rightline{
\rightline{\includegraphics[width=0.25\linewidth]{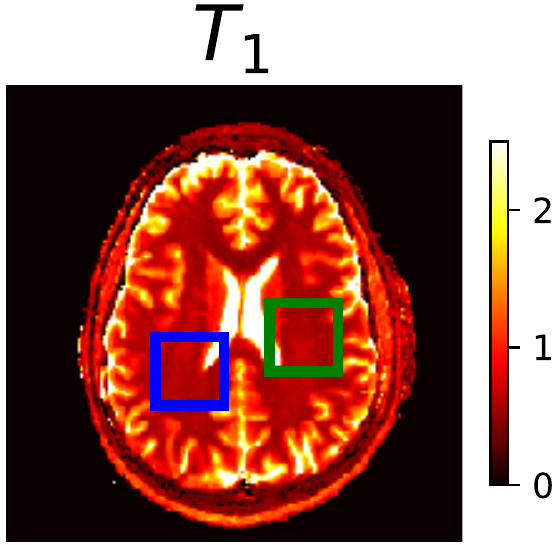}}
\rightline{\includegraphics[width=0.25\linewidth]{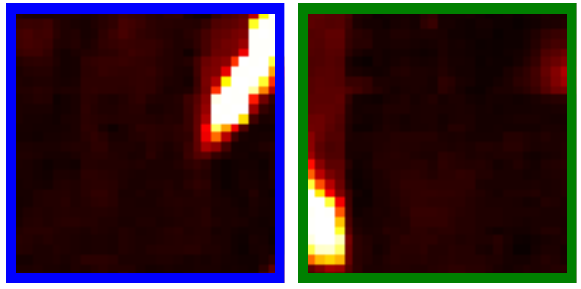}}
%}
\includegraphics[width=1\linewidth]{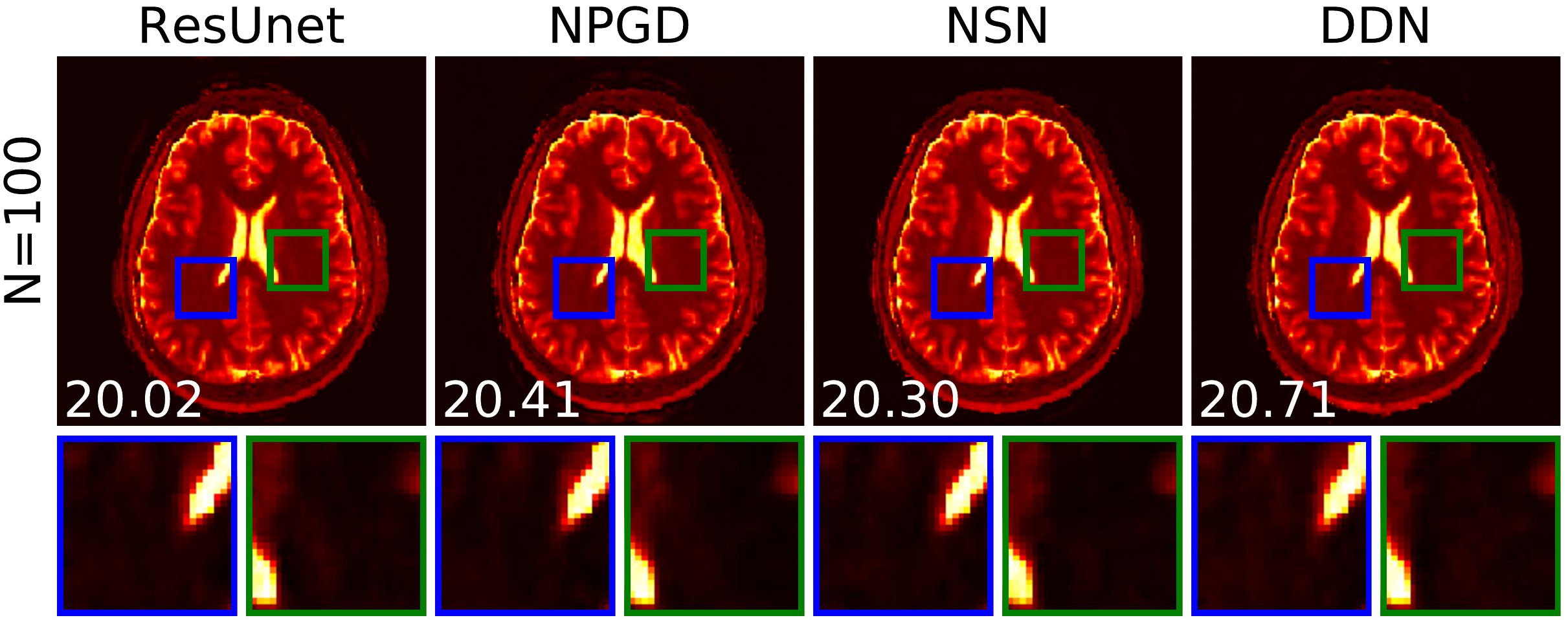}
\includegraphics[width=1\textwidth]{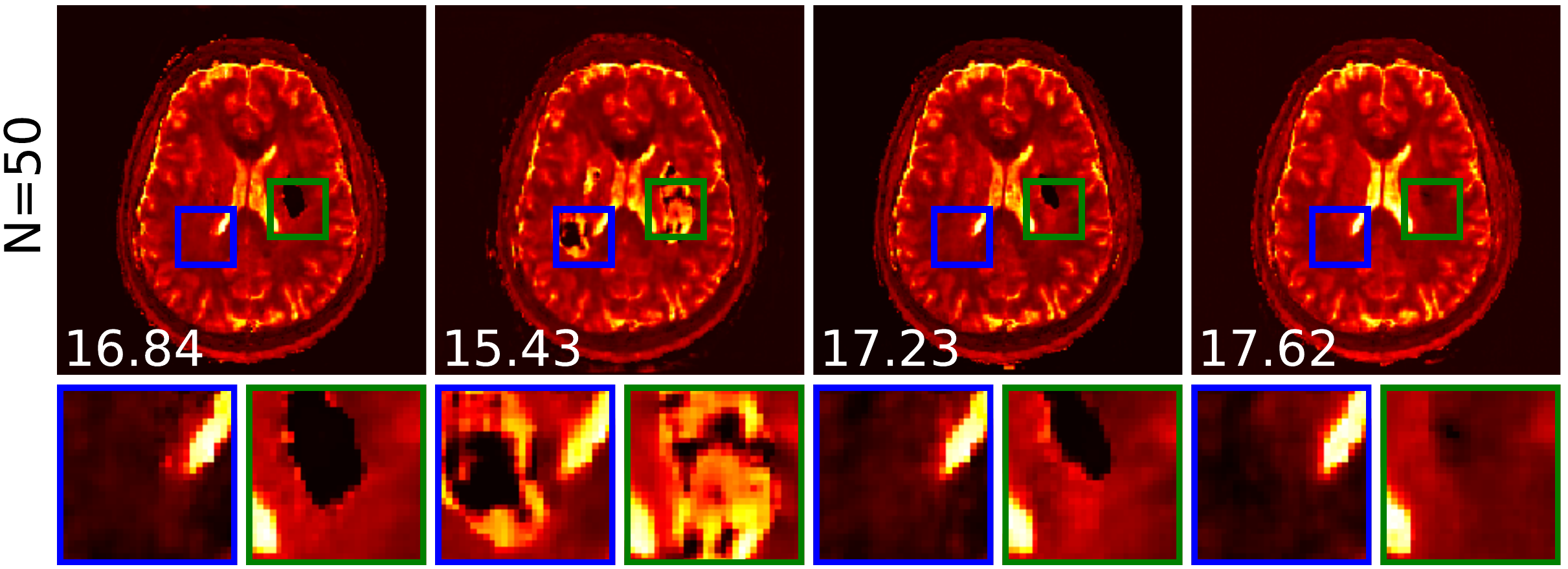}
\includegraphics[width=1\textwidth]{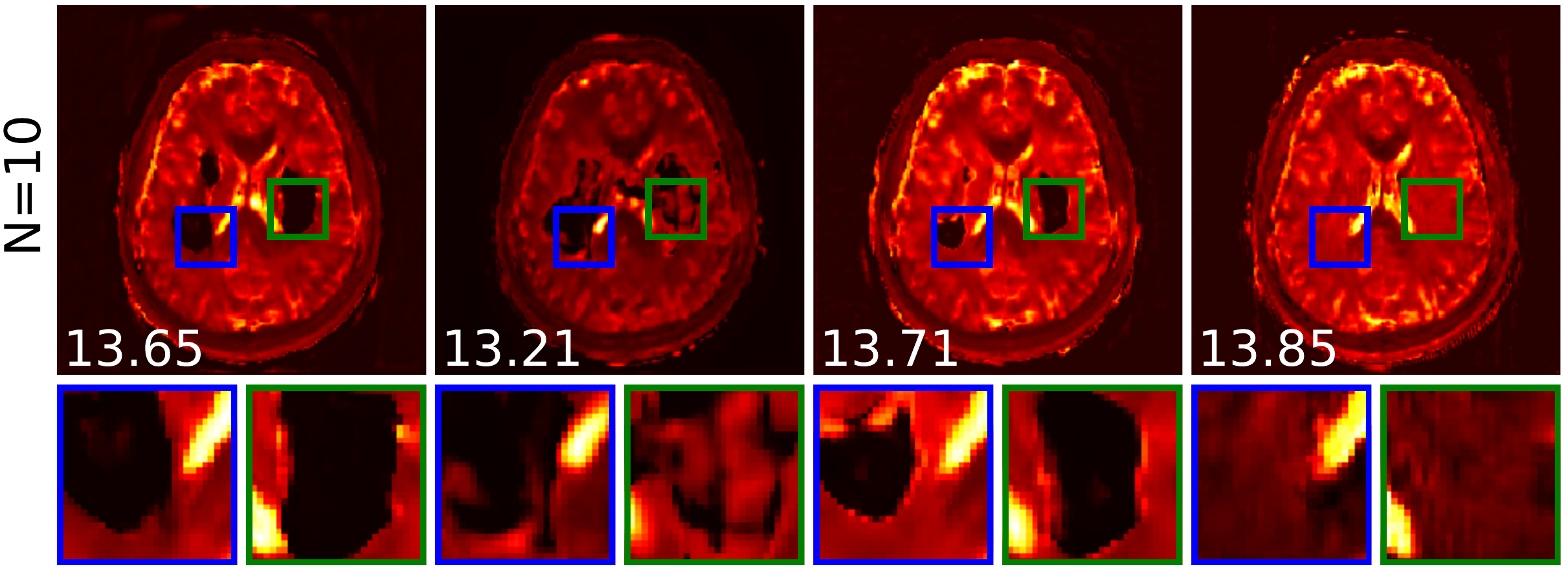}
  \caption{prediction of $\emm$.}
  \label{fig:reconstruction_T1}
\end{subfigure}
\caption{(a) Reconstruction of MRF image $\ex$ from its noisy measurement $\ey_{\epsilon}$ ($\sigma_{\epsilon}=0.01$) and (b) its corresponding $T_1$ map prediction. From top to bottom are the groundtruth x and m and their corresponding reconstructions by ResUnet \cite{jin2017deep}, NPGD\cite{mardani2018neural}, NSN\cite{schwab2018deepnull} and DDN (proposed) trained with 100, 50 and 10 sample, respectively.}
%\caption{(a).}
\end{figure*}

Table \ref{table:tab_csmrf} reports the inferring time (IT), NMSE and GE for the reconstruction of MRF images and the PSNRs for the tissue maps prediction results. Due to space limitation, we only list the PSNR results of $ T_1 $ here. Fig.\ref{fig:reconstruction_x} and Fig.\ref{fig:reconstruction_T1} show the reconstruction of a test MRF image $\ex$'s first channel and $\ex$'s $T_1$ map prediction, respectively. In summary, we make three key findings.

First, DDN obtained the best reconstruction results and the best generalization ability in all cases (Table \ref{table:tab_csmrf}). When training with fewer training samples \egB 50 and 10, NSN and NPGD perform better than NPGD. This demonstrates that a neural network that improperly-trained using a few samples is not suitable as a proximal operator. ResUnet is more stable than NPGD because the former does not need to be used in another optimization algorithm which is irrelevant to the neural network training.

Second, let  $\tau$ denotes the number of times a particular model accesses the physical model (\egB $\eH$, $\eHT$). Clearly, $\tau_{\text{ResUnet}}=0$, $\tau_{\text{NSN}}=2$,  $\tau_{\text{DDN}}=2\times \tau_{\text{NSN}}=4$ and $\tau_{\text{NPGD}}=2\times3=6$ since NPGD needs to access both $\eH$ and $\eH^{\top}$ in a proximal gradient descent iteration. As can be seen, the inferring time of ResUnet is very short due to its essence of the pure physics-free model. Although NPGD requires longer inferring time ($2.5$ times DDN), it only performs well when training using 100 samples, \ieB fewer samples are not sufficient to train a good neural proximal operator \cite{rick2017one,mardani2018neural}. This also shows the usage of physics $\PP_r$ and $\PP_n$ in the proposed decomposition learning is reasonable since it obtained best results within an acceptable inferring time.

%Finally, there exists lots of artifacts in the reconstruction (Fig.\ref{fig:reconstruction_x} and Fig.\ref{fig:reconstruction_T1})  by ResUnet, NPGD and NSN  when training them with 50 or 10 samples.

Finally, when the ResUnet, NPGD and NSN are trained with 50 or 10 samples, there exist lots of artifacts in their reconstructions (see Fig.\ref{fig:reconstruction_x} and Fig.\ref{fig:reconstruction_T1}). In contrast, DDN can stably output reasonable MRF image reconstructions and such that lead to a physics plausible prediction of $T_1$ map. This shows the use of decomposition learning can help improve the generalization ability of the neural network and alleviate the dependence of deep learning performance on large-volume training samples.

\subsection{Ablation study}
We are interested in studying (1) the impact of different training strategies (jointly/decoupled) and different connection types (independent/cascade) to DDN and (2) the importance of component $\PP_r$ and $\PP_n$ to DDN. To do that, we consider the Super-resolution (SR) task and train the models on the BSDS300 \cite{agustsson2017ntire} and test on the benchmark dataset Set5\cite{bevilacqua2012low}. We super-resolve low-resolution (LR) images by a scale factor of 2. The LR images are formed by downsampling with bi-cubic anti-aliasing the HR images \cite{dong2015image}, and then corrupting them with Gaussian noise with a standard deviation $\sigma_{\epsilon}=0.1$ ($10\%$ noise). The presence of anti-aliasing in the downsampling means that the bi-cubic upsampling operator is a reasonable approximation to $\eHT$, and this is what we use in our DDN. Training is performed using $40\times 40$ RGB image patches (LR). The quantitative results presented are then evaluated on the luminance (Y) channel. We \emph{emphasize} that our goal here is not to achieve state-of-the-art performance on the SR task, but a simple scenario to study the behaviour of ablation analysis for decomposition learning.

\subsubsection{Training strategy study} We trained DDN using different training strategies. The test results are reported in the Table \ref{tab:strategy} which demonstrates the cascade architecture always performs slightly better than the independent one in term of both PSNR and GE. Note that for the independent model the joint and decoupled training are exactly equivalent. We also find that $\F$ and $\G$ benefit from more iterations in the decoupled training, especially for very noisy cases. In contrast, joint training was observed to be more efficient. In all experiments, we therefore use the cascade architecture with joint training for the DDNs.

\begin{table}[t]
\caption{Comparison results of ablation study on image SR task ($\times2$, $\sigma_{\epsilon}=0.1$).}
  \centering
  \fontsize{8}{8}\selectfont %\fontsize{6.5}{5.3}{7.0}{6.0}
  \begin{threeparttable}
\begin{tabular}{c | c | c| c |c |c |c |c}
\toprule
 \multirow{3}*{Metric}& \multicolumn{3}{c|}{Training strategy study}  & \multicolumn{4}{c}{Ablation study} \\ \cline{2-8}
 & Independent  & \multicolumn{2}{c|}{Cascade} & \multirow{2}*{$\text{DDN}_1$} & \multirow{2}*{$\text{DDN}_2$}& \multirow{2}*{$\text{DDN}_3$} & \multirow{2}*{$\text{DDN}_4$} \\ \cline{2-4}
  & Joint/Decoupled  & Joint & Decoupled &  & & &  \\ \midrule
PSNR (dB)   & 26.15 & 26.30 &\textcolor{blue}{26.38} & 22.31 & 20.95 & 25.82 & \textcolor{blue}{26.30}\\\midrule
GE ($10^{-3}$) & \textcolor{blue}{1.24} &1.26   &1.24&1.75 & 1.41 &1.83 &\textcolor{blue}{1.26}\\
\bottomrule
\end{tabular}
\end{threeparttable}
\label{tab:strategy}
\end{table}

\begin{figure}[ht]
\centering
\includegraphics[width=1\linewidth]{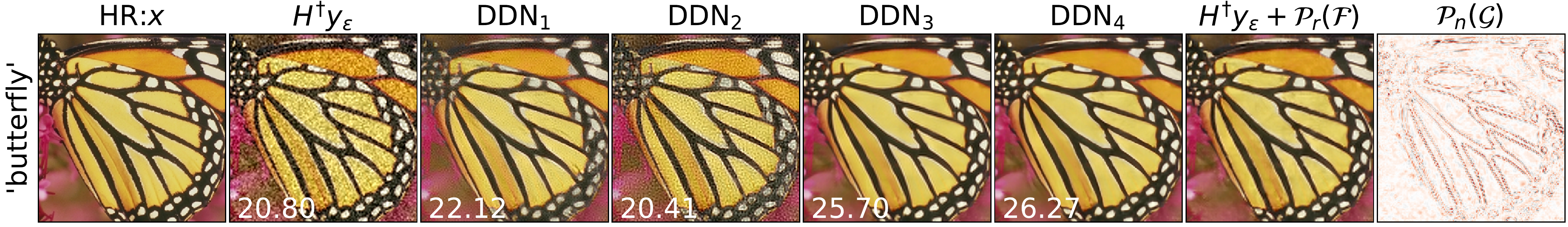}
\includegraphics[width=1\textwidth]{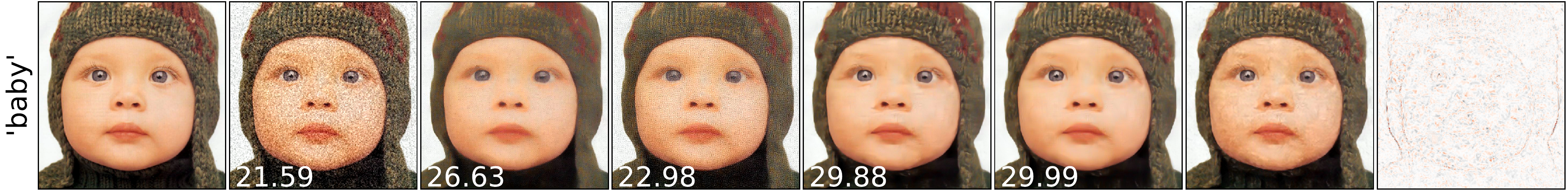}
\caption{Visualization of DDN with different specs on the  SR task ($\times2$, $\sigma_{\epsilon}=0.1$). From the left to right are the oracle HR image $\ex$, reconstructions by $\eHT\ey_{\epsilon}$, $\text{DDN}_1$(\chkNO$\PP_r$, \chkNO$\PP_n$), $\text{DDN}_2$(\chkNO$\PP_r$, \chkYES$\PP_n$), $\text{DDN}_3$(\chkYES$\PP_r$, \chkNO$\PP_n$), $\text{DDN}_4$(\chkYES$\PP_r$, \chkYES$\PP_n$), and the $r$-component $
\eHT\ey_{\epsilon}+\PP_r(\F)$ and $n$-component $\PP_n(\G)$ recovered by $\text{DDN}_4$.}
\label{fig:sr}
\end{figure}

\subsubsection{Importance of $\PP_r$ and $\PP_n$}
%To verify the importance of the component $\PP_r$ and $\PP_n$ in DDN,
We trained four DDN variants by different usages of $\PP_r$ and $\PP_n$: $\text{DDN}_1$ - deactivate both $\PP_r$ and $\PP_n$, $\text{DDN}_2$ - only use $\PP_n$, $\text{DDN}_3$ - only use $\PP_r$ and $\text{DDN}_4$ - use both $\PP_r$ and $\PP_n$ . The statistics PSNRs and GEs of the reconstruction are listed in Table \ref{tab:strategy}. Two reconstruction examples and the learnt $r$-component and $n$-component by $\text{DDN}_4$ are shown in Fig.\ref{fig:sr}. From the results, we have the following conclusions.

First, it can be seen that the $\text{DDN}_4$ achieves the best results which used both projectors $\PP_r$ and $\PP_n$. $\text{DDN}_2$ is essentially a nullspace network and performed poorly in the presence of noise, as in this scenario the denoising task plays a significant role in the inverse problem. $\text{DDN}_1$ is equivalent to residual learning and is purely learning-based and physics-free, it consistently provides better performance than $\text{DDN}_2$, but it has the higher GEs. This suggests that introducing physics into deep learning models can facilitate the neural network to enjoy better generalization for solving inverse problems.

Second, $\text{DDN}_3$ performs better than $\text{DDN}_1$ and $\text{DDN}_2$ but worse than $\text{DDN}_4$. It is because $\text{DDN}_3$ considers denoising by recovering the $r$-component of data but fails to learn the $n$-component. This demonstrates integrating $\PP_r$ and $\PP_n$ into the proposed decomposition learning allows the DDN to remove structural noise and to simultaneously accurately approximate the $r$-component $\PP_r(\ex)$  (see the penultimate column in Fig.\ref{fig:sr}) and predict the $n$-component $\PP_n(\ex)$ (see the last column in Fig.\ref{fig:sr}) from its noisy measurement. From the above discussions, we conclude that decomposition learning is well-principled, structurally simple, and highly interpretable.

%Note that theoretically, with a sufficiently rich architecture and enough training data, $\text{DDN}_1$ should be able to do well even though it is physics free.

Finally, we emphasize that given a linear forward operator $\eH$, the decomposition learning naturally exists and is easy to define. One can plug the decomposition learning, into other existing specialized deep learning solvers for different inverse problems,  with which we believe one could increase the performance limit of the deep learning solvers. %However, we leave this as our future work.

\section{Conclusion}
In this paper, we have proposed a deep decomposition learning framework for building an end-to-end and physics-engaged neural network solution for inverse problems. We have explicitly reformulated the neural network layers to learn range-nullspace decomposition functions with reference to the layer inputs, instead of learning unreferenced functions. We have shown that the decomposition networks not only produce superior results, but also enjoy good interpretability and generalization. We have demonstrated the advantages of decomposition learning on CS-MRF and image super-resolution examples. In future work, we will explore adapting the proposed deep decomposition learning to more challenging inverse problems such as tomographic imaging. %This will require the investigation of how best (and how accurately) %to approximate the pseudo-inverse $\eHT$ and the associated projection operators.

\section*{Acknowledgements}
%\noindent\textbf{Acknowledgements.}
We thank Pedro G\'{o}mez, Carolin Prikl and Marion Menzel from GE Healthcare and Mohammad Golbabaee from Bath University for the quantitative anatomical maps dataset. DC and MD are supported by the ERC C-SENSE project (ERCADG-2015-694888).

\clearpage

\bibliographystyle{splncs04}
\bibliography{egbib}
\end{document}